\documentclass[twocolumn,pra,showpacs,superscriptaddress,floatfix]{revtex4}

\usepackage{graphicx}
\usepackage[latin1]{inputenc}
\usepackage{graphicx,epstopdf}
\usepackage{dcolumn}
\usepackage{amsmath}
\usepackage[hypertex]{hyperref}
\usepackage{amsfonts}
\usepackage{amsthm}
\usepackage{amssymb}
\usepackage{mathrsfs}
\PassOptionsToPackage{caption=false}{subfig}
\usepackage{subfig}
\usepackage{color}
\usepackage[normalem]{ulem}

\begin{document}

\newcommand{\red}[1]{\textcolor{red}{#1}}
\newcommand{\blue}[1]{\textcolor{blue}{#1}}

\newcommand{\inprod}[2]{\braket{#1}{#2}}
\newcommand{\Tr}{\mathrm{Tr}}
\newcommand{\vp}{\vec{p}}
\newcommand{\Or}{\mathcal{O}}
\newcommand{\so}[1]{{\ignore{#1}}}

\title{Geometric quantum discord through the Schatten 1-norm}

\author{F. M. Paula}
\email{fagner@if.uff.br}

\affiliation{Instituto de F\'isica, Universidade Federal Fluminense, Avenida Gal. Milton Tavares de Souza s/n, Gragoat\'a, 
24210-346, Niter\'oi, RJ, Brazil}

\author{Thiago R. de Oliveira}

\affiliation{Instituto de F\'isica, Universidade Federal Fluminense, Avenida Gal. Milton Tavares de Souza s/n, Gragoat\'a, 
24210-346, Niter\'oi, RJ, Brazil}

\author{M. S. Sarandy}

\affiliation{Instituto de F\'isica, Universidade Federal Fluminense, Avenida Gal. Milton Tavares de Souza s/n, Gragoat\'a, 
24210-346, Niter\'oi, RJ, Brazil}

\date{\today}

\begin{abstract}

It has recently been pointed out that the geometric quantum discord, as defined by the Hilbert-Schmidt norm ($2$-norm), 
is not a good measure of quantum correlations, since it may increase under local reversible operations on the unmeasured 
subsystem. Here, we revisit the geometric discord by considering general Schatten $p$-norms, explicitly showing that the 
1-norm is the only $p$-norm able to define a consistent quantum correlation measure. In addition, by restricting 
the optimization to the tetrahedron of two-qubit Bell-diagonal states, we provide an analytical expression for 
the 1-norm geometric discord, which turns out to be equivalent to the negativity of quantumness. We illustrate the 
measure by analyzing its monotonicity properties.
\end{abstract}

\pacs{03.65.Ud, 03.67.Mn, 75.10.Jm}

\maketitle

Quantum discord is an information-theoretic measure of non-classical correlations, initially proposed by Ollivier 
and Zurek~\cite{Ollivier}, which goes beyond entanglement (i.e., separable states can have nonzero discord) and whose 
characterization has attracted much attention during the last decade (see Ref.~\cite{Modi} 
for a review and Ref.~\cite{Gu:12} for an operational interpretation). From an 
analytical point of view, the evaluation of quantum discord is a difficult task, even for (general) two-qubit states, 
since an optimization procedure is required for the conditional entropy over all local generalized measurements. 
In this scenario, closed expressions are known only for classes of states~\cite{Luo,Ali}.

The difficulty of extracting analytical solutions for quantum discord led Daki\'c, Vedral, and Brukner to propose a geometric 
measure of quantum discord \cite{Dakic}, which quantifies the amount of quantum correlations of a state in terms of its 
minimal Hilbert-Schmidt distance from the set of classical states. The calculation of this alternative measure requires a 
simpler minimization process, which is realizable analytically for general two-qubit states \cite{Dakic} as well as for 
arbitrary bipartite states \cite{Lu-Fu,Hassan,Rana}. Moreover, it has been shown to exhibit operational significance in specific 
quantum protocols (see, e.g., Ref.~\cite{Dakic:12}). Despite those
remarkable features, geometric discord is known to be sensitive to
the choice of distance measures (see, e.g., Ref.~\cite{Bellomo}).
In turn, as recently pointed out~\cite{Hu:12,Tufarelli:12,Piani}, 
the geometric discord as proposed in Ref.~\cite{Dakic} cannot be regarded as a good measure for the quantumness of 
correlations, since it may increase under local operations on the unmeasured subsystem. In particular, it has explicitly been 
shown by Piani~\cite{Piani} that the simple introduction of a factorized local ancillary state on the unmeasured party changes the geometric discord by a factor given by the lack of purity of the ancilla. This is in contrast with the entropic quantum 
discord, which does not suffer this problem. From a technical point of view, the root of this drawback is the lack of contractivity of geometric discord under trace-preserving quantum channels. Remarkably, this is strongly connected with the norm adopted to define distance in the state space.

Most recently, Tufarelli {\it et al.}~\cite{Tufarelli:13} have introduced a modified version of geometric discord that is immune to the particular ancilla considered in Ref.~\cite{Piani}. However, since this measure is also based on Hilbert-Schmidt 
distance, it inherits the noncontractivity problem (see, e.g., examples in Ref.~\cite{Hu:12}). A way to circumvent this issue  is to employ the trace distance in place of the Hilbert-Schmidt norm~\cite{Hu:12,Nielsen-Chuang,Rana:13}.
In this direction, we consider the generalization of the geometric discord in terms of Schatten $p$-norms. More specifically, we show that the geometric discord as defined by the 1-norm is the only $p$-norm geometric discord invariant under the class of channels considered in Ref.~\cite{Piani}. Furthermore, by restricting the minimization to states in the Bell-diagonal form, we analytically evaluate the 1-norm geometric discord for arbitrary Bell-diagonal two-qubit states. As an illustration, we compare our result with the entropic quantum discord and the 2-norm geometric discord, analyzing its monotonicity properties as a function of the correlation functions.

{\it Entropic and geometric measures of quantum discord}. Quantum discord has been introduced as an entropic measure of quantum correlation 
in a quantum state. For a bipartite system described by the 
density matrix $\rho$, it is defined by the difference 
${\cal Q}(\rho)=\mathcal{I}(\rho)-\mathcal{J}(\rho)$~\cite{Ollivier},
where $\mathcal{I}(\rho)$ is the quantum mutual information, which represents the total correlation in $\rho$~\cite{Groisman:05}, and $\mathcal{J}(\rho)$ is the measurement-based mutual information, which can be interpreted as the classical correlation in $\rho$~\cite{Henderson:01}.   
These quantities are given by $\mathcal{I}(\rho)=S(\rho_{a})+S(\rho_{b})-S(\rho)$ and $\mathcal{J}(\rho)=S(\rho_{b})- \min_{\{E_{k}\}}\left [\sum_{k}p_{k}S(\rho_{b|k})\right ]$. In these expressions, $S(\rho)=-\text{tr}\left[\rho\log_{2}\rho\right ]$ denotes the von Neumann entropy, 
$\rho_{a(b)}$ is the reduced density matrix of the subsystem $a(b)$, and the minimum is taken over all 
possible positive operator-valued measures (POVMs) $\{E_{k}\}$ on subsystem $a$, where 
$\rho_{b|k}=\text{tr}_{a}\left [E_{k}\rho \right ]/p_{k}$ is the post-measurement state of $b$ after the outcome $k$ on $a$ is obtained with probability $p_{k}=\text{tr}\left [E_{k}\rho\right ]$. 
 
\begin{figure}[h]
\includegraphics[scale=0.27]{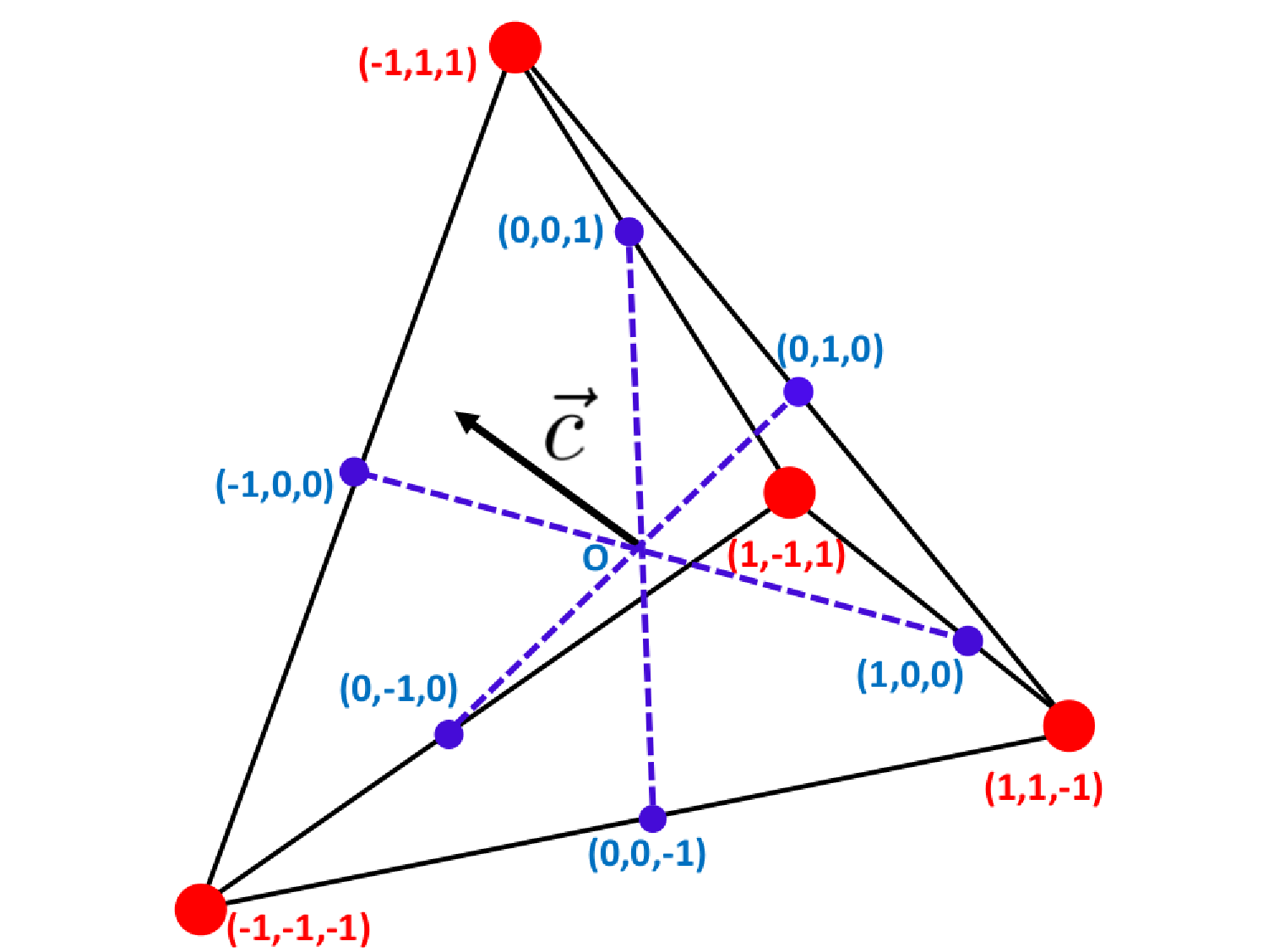} \caption{\label{fig:tetraedro} (Color online) Tetrahedron corresponding to the two-qubit Bell-diagonal states, with its vertices representing the four Bell states. Quantum discord is a maximum (${\cal Q}=1$ and $D_{G}=1/2$) in these vertices and vanishing (${\cal Q}=D_{G}=0$) over the perpendicular axes $c_{1},c_{2}$, and $c_{3}$ (dashed lines).}
\end{figure}

The analytical minimization over POVMs involved in $\mathcal{J}(\rho)$ constitutes a hard task, even for two-qubit 
systems in a general state. This motivated the introduction of an alternative measure~\cite{Dakic}, which was named 
geometric quantum discord. Such a geometric measure is based on the distance between the given quantum state $\rho$ 
and the closest classical-quantum state $\rho_{c}$, reading
\begin{equation}\label{eq:DG}
D_{G}(\rho)=\min_{\Omega_0}\left\Vert \rho-\rho_{c}\right\Vert^2_{2},
\end{equation}
where $\left\Vert X\right\Vert_{2}=\sqrt{\text{tr}\left[X^{\dagger}X\right]}$ is the Hilbert-Schmidt norm ($2$-norm)   
and $\Omega_0$ is the set of classical-quantum states, whose general form is given by
\begin{equation}
\rho_{c}=\sum_{k}p_{k}\Pi^a_{k}\otimes\rho^{b}_{k} ,
\label{rho-classical}
\end{equation}
with $0 \le p_k \le 1$ ($\sum_{k} p_k = 1$), $\{\Pi^a_{k}\}$ denoting a set of orthogonal projectors for subsystem $a$, 
and $\rho^{b}_{k}$ being a general reduced density operator for subsystem $b$. Note that extremization here is over a 
distance measure rather than POVMs, as in $\mathcal{J}(\rho)$. In terms of the entropic quantum discord ${\cal Q}(\rho)$ and of the negativity (of entanglement) ${\cal N}(\rho)=\left\Vert \rho^{t_{a}}\right\Vert_{1}-1$, where $\rho^{t_{a}}$ denotes partial transposition of $\rho$ with respect to
subsystem $a$ and $\left\Vert X\right\Vert_{1}=\text{tr}\left[\sqrt{X^{\dagger}X}\right]$ is the trace norm, the geometric discord  presents the 
following bound for two-qubit states~\cite{Girolami}:
\begin{equation}
2 D_{G}(\rho) \geq {\cal Q}^2(\rho),{\cal N}^2(\rho). \label{eq:INEQ} 
\end{equation}
The inequality $2D_{G}\geq {\cal N}^2$ is not universal, with counterexamples in spaces of dimension higher than 
$2\times2$~\cite{Rana:R}.

We will focus here in the particular case of two-qubit Bell diagonal states, whose density operator presents the form
\begin{equation}\label{bell}
\rho=\frac{1}{4}\left[I\otimes I+\vec{c}\cdot \left(\vec{\sigma}\otimes\vec{\sigma}\right)\right],
\end{equation}
where  $I$ is the identity matrix, $\vec{c}=\left(c_{1},c_{2},c_{3}\right)$ is a three-dimensional vector  and 
$\vec{\sigma}=\left(\sigma_{1},\sigma_{2},\sigma_{3}\right)$ is a vector formed by Pauli matrices. In this case, the entropic quantum discord and the geometric discord are given by \cite{Luo,Lu}
\begin{equation}\label{D}
{\cal Q}=\log_{2}\frac{4\lambda_{00}^{\lambda_{00}}\lambda_{01}^{\lambda_{01}}\lambda_{10}^{\lambda_{10}}\lambda_{11}^{\lambda_{11}}}{(1-c_{+})^{\frac{1-c_{+}}{2}}(1+c_{+})^{\frac{1+c_{+}}{2}}}
\end{equation} 
and
\begin{equation}\label{D2}
D_{G}=\frac{1}{4}\left(c_{-}^{2}+c_{0}^{2}\right),
\end{equation}
where $\lambda_{ij}=\left[1+(-1)^{i}c_{1}-(-1)^{i+j}c_{2}+(-1)^{j}c_{3}\right]/4$ are the eigenvalues of the density operator $\rho$, whereas $c_{+}=\text{max}\left[\left|c_{1}\right|,\left|c_{2}\right|,\left|c_{3}\right|\right]$, $c_{0}=\text{int}\left[\left|c_{1}\right|,\left|c_{2}\right|,\left|c_{3}\right|\right]$, and $c_{-}=\text{min}\left[\left|c_{1}\right|,\left|c_{2}\right|,\left|c_{3}\right|\right]$ represent the maximum, intermediate, and minimum among the absolute values of the correlation functions $c_{1}$, $c_{2}$, and $c_{3}$, respectively. If $\rho$ describes a physical state, then $0\leq \lambda_{ij}\leq 1$ and $\sum_{i,j} \lambda_{ij}=1$. In this condition, the vector  $\vec{c}$ must be restricted to the tetrahedron whose vertices situated on the points $(1,1,-1)$, $(-1,-1,-1)$, $(1,-1,1)$, and $(-1,1,1)$ represent the Bell states 
(see Fig.~\ref{fig:tetraedro}). Quantum discord is a maximum (${\cal Q}=1$ and $D_{G}=1/2$) in these 
vertices and minimum (${\cal Q}=D_{G}=0$) over the perpendicular axis $c_{1},c_{2}$, and 
$c_{3}$ (dashed lines). 

{\it Geometric quantum discord and Schatten p-norms}. Despite being easier to compute and exhibiting an interesting geometric interpretation, 
the measure $D_G$ fails as a rigorous quantifier of quantum correlation, since it may 
increase under local reversible operations on the unmeasured subsystem. Explicitly, 
by assuming the map $\Gamma^{\sigma}:X\rightarrow X\otimes\sigma$, i.e., a 
channel that introduces a noisy ancillary state, Piani has recently shown that \cite{Piani} $D_{G}(\Gamma^{\sigma}_b\left[\rho\right])=D_{G}(\rho)\text{tr}\left[\sigma^{2}\right]$. This means that the geometric discord may increase under local operations on the unmeasured subsystem 
$b$, because $\text{tr}\left[\sigma^{2}\right]\leq 1$ in general. Indeed, by considering the coupling 
of $b$ with an arbitrary auxiliary system in a mixed state $\sigma$, we obtain that $D_{G}$ increases by the 
simple reversible removal of $\sigma$. The origin of this problem is the Hilbert-Schmidt norm, which is not 
an appropriate choice for geometrically quantifying the quantumness of correlations (for a similar analysis in the 
case of entanglement, see Ref.~\cite{Ozawa:00}).

Let us then consider the geometric discord based on a more general norm, defined by \cite{Debarba}
\begin{equation}
D_{p}(\rho)=\min_{\Omega_0}\left\Vert \rho-\rho_{c}\right\Vert^p_{p},
\end{equation}
where $\left\Vert X\right\Vert_{p}=\text{tr}\left[\left(X^{\dagger}X\right)^{\frac{p}{2}}\right]^{\frac{1}{p}}$ is the Schatten $p$-norm, with $p$ denoting a positive integer number. In this notation, the geometric discord is simply obtained by taking $p=2$, namely, 
$D_G = D_2$. Since the $p$-norm is multiplicative under tensor products \cite{Aubrun}, it is then easy to see that $\left\Vert X\right\Vert_{p}\rightarrow\left\Vert \Gamma^{\sigma}_{b}\left[X\right]\right\Vert_{p}=\left\Vert X\right\Vert_{p}\left\Vert \sigma \right\Vert_{p}$. Thus,
\begin{equation}
D_{p}(\Gamma^{\sigma}_{b}\left[\rho\right])=D_{p}(\rho)\left\Vert \sigma \right\Vert_{p}^{p}.
\end{equation}
Note that $\left\Vert \sigma \right\Vert_{p}=1$ if and only if $p=1$, since 
$\left\Vert \sigma \right\Vert_{1}=\text{tr}\left[\sigma\right]=1$ for a general state $\sigma$. 
Therefore,  the geometric discord based on the 1-norm is the only possible Schatten $p$-norm able to consistently quantify non-classical correlations. Indeed, one can show that $D_{1}(\rho)$ is non-increasing under general local operations on $b$ 
(see also Ref.~\cite{Hu:12}). Due to the properties of the trace distance, the 1-norm geometric discord 
is contractive under trace-preserving quantum channels~\cite{Hu:12,Nielsen-Chuang}, i.e. $\left\Vert \rho-\rho_{c}\right\Vert_{1} \ge \left\Vert \varepsilon(\rho)-\varepsilon(\rho_{c})\right\Vert_{1}$, where $\varepsilon$ is a general trace-preserving quantum operation. Then, let us consider a quantum operation 
$\varepsilon_b$, which acts only over subsystem $b$. By denoting as $\overline{\rho}_c$ the closest classical state 
to a given quantum state $\rho$, we can write $D_1(\rho) =  \left\Vert \rho-\overline{\rho}_{c}\right\Vert_{1} 
\ge \left\Vert \varepsilon_b(\rho)-\varepsilon_b(\overline{\rho}_{c})\right\Vert_{1}$. Note that $\varepsilon_b(\overline{\rho}_{c})$ is still a classical state, but it is not {\it necessarily} the closest classical 
state to $\varepsilon_b(\rho)$. Then, 
$\left\Vert \varepsilon_b(\rho)-\varepsilon_b(\overline{\rho}_{c})\right\Vert_{1} \ge D_1(\varepsilon_b(\rho))$. Hence it follows 
that $D_1(\rho) \ge D_1(\varepsilon_b(\rho))$~\cite{Hu:12}, 
which implies that $D_1(\rho)$ cannot increase under operations over subsystem $b$. 

{\it 1-norm geometric quantum discord for Bell-diagonal states}. In order to obtain the 1-norm geometric discord for two-qubit systems described 
by Bell-diagonal states given by Eq.~(\ref{bell}), 
let us start from the expression
\begin{equation}\label{D1}
D_{1}(\rho)=\min_{\Omega_0}\left\Vert \rho-\rho_{c}\right\Vert_{1},
\end{equation}
where $\left\Vert X\right\Vert_{1}=\text{tr}\left[\sqrt{X^{\dagger}X}\right]$ is the $1$-norm, $\rho$ is given 
by Eq. (\ref{bell}) and $\rho_{c}$ is an arbitrary classical-quantum state given by Eq.~(\ref{rho-classical}). 
The minimization over the whole set of classical states was obtained for the 2-norm \cite{Lu-Fu} and the relative entropy \cite{Modi:10}, where it can be proved that the minimal
state is the measured original state. We will make
a similar hypothesis and assume that the minimal state preserves the Bell-diagonal form of the original state. This has been numerically checked for a number of Bell-diagonal states, as will be discussed below. Therefore, we assume that the minimization in Eq.~(\ref{D1}) is achieved by a Bell-diagonal 
classical state $\rho_{c}^{(BD)}$, which is denoted by
\begin{equation}
\rho_{c}^{(BD)} = \frac{1}{4}\left[I\otimes I+\vec{l} \cdot \left(\vec{\sigma}\otimes\vec{\sigma}\right)\right], 
\label{rhoc-BD}
\end{equation}
with $\vec{l}$ representing a vector over the perpendicular classical axes in the tetrahedron of Bell-diagonal states 
(dashed lines in Fig.~\ref{fig:tetraedro}). Then, $\vec{l}$ has the form $\vec{l}_{1}=(l_1,0,0)$, $\vec{l}_{2}=(0,l_2,0)$, or $\vec{l}_{3}=(0,0,l_3)$, with 
$l_i \in \Re$ and $-1 \le l_i \le 1$ . From Eqs.~(\ref{D1}) and (\ref{rhoc-BD}), we can then write    
\begin{equation}
\label{eq:dmin}
D_{1}= \min\left[\min_{l_{1}} f_{1}(l_{1}),\min_{l_{2}} f_{2}(l_{2}),\min_{l_{3}} f_{3}(l_{3})\right],
\end{equation}
where 
\begin{equation}\label{eq:min-0}
f_{i}(l_{i})=\left\Vert\frac{1}{4}(\vec{c} - \vec{l}_{i}) \cdot \left(\vec{\sigma}\otimes\vec{\sigma}\right)\right\Vert_{1}=\sum_{p=0}^{1}\sum_{q=0}^{1}\left|\tau_{pq,i}\right|
\end{equation}
with $\tau_{pq,i}=\left[(-1)^{p}(c_{i}-l_{i})-(-1)^{p+q}c_{j}+(-1)^{q}c_{k}\right]/4$ $(i\neq j \neq k)$ denoting the eigenvalues of the operator $(\vec{c} - \vec{l}_{i}) \cdot \left(\vec{\sigma}\otimes\vec{\sigma}\right)/4$. Now, by defining $d_{i}=l_{i}-c_{i}$ and $d_{\pm}=c_{k}\pm c_{j}$, we find $f_{i}(d_{i})=\left(\left|d_{i}+d_{+}\right|+\left|d_{i}-d_{+}\right|+\left|d_{i}+d_{-}\right|+\left|d_{i}-d_{-}\right|\right)/4$. Because $f(d_{i})$ reaches its minimum value when $d_{i}=0$, then $\min_{l_{i}} f_{i}(l_{i})=\min_{d_{i}} f_{i}(d_{i})=\max\left[\left|c_{j}\right|,\left|c_{k}\right|\right]$. By using this result in Eq.~(\ref{eq:dmin}), we then obtain
\begin{equation}
D_{1} =c_{0} = \text{int}\left[\left|c_{1}\right|,\left|c_{2}\right|,\left|c_{3}\right|\right].
\label{1normGD}
\end{equation}
The same result encapsulated by Eq.~(\ref{1normGD}) was obtained in the context of the study of the negativity of quantumness, which is a measure of nonclassicality 
recently introduced in Refs.~\cite{Piani:NoQ,Nakano:12} and experimentally discussed  in Ref.~\cite{Silva:12}. 
In such a case, the 1-norm distance is computed with respect to the decohered (measured) state $\rho^{,}=\sum_{k}\Pi^a_{k}\rho \Pi^a_{k}$.

\begin{figure}[ht!]
\includegraphics[scale=0.4]{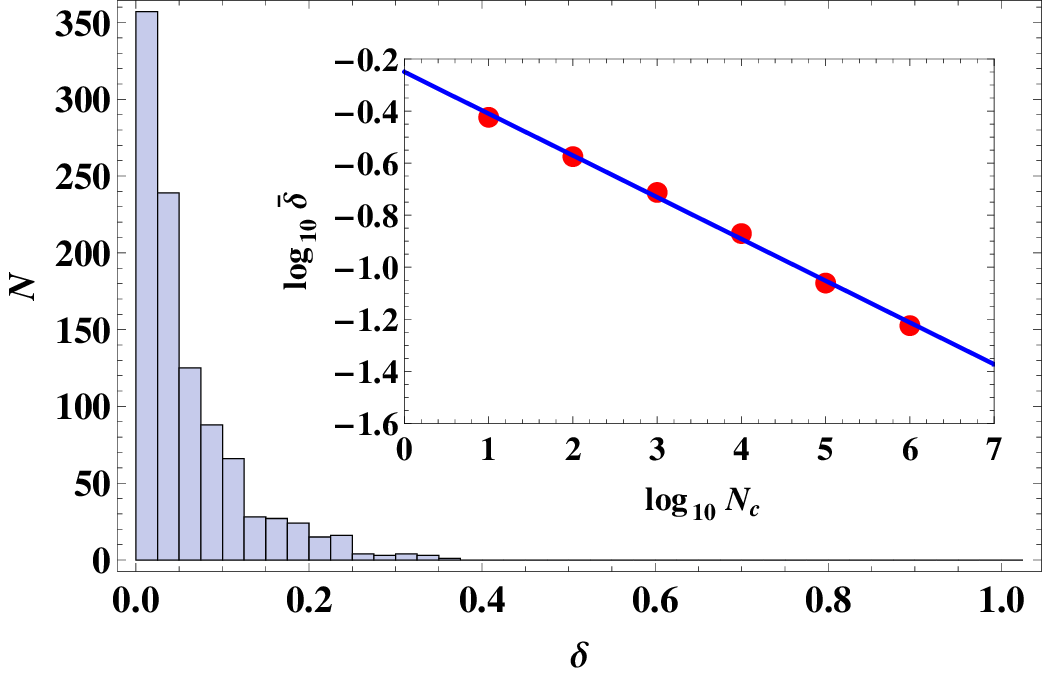} \caption{\label{fig:fignum} (Color online) Histogram of  $\delta$ for $N=10^3$ 
Bell-diagonal states and $N_{c} = 10^6$ classical states. In the inset, we show the decreasing behavior of 
$\log_{10}\bar{\delta}\times\log_{10}N_{c}$ for $N_{c}=10$, $10^2,$ $10^3,$ $10^4,$ $10^5,$ and $10^6$.}
\end{figure}

For a finite subset of classical states $\Omega_0^\prime$, the equivalence between 
Eqs.~(\ref{D1}) and (\ref{1normGD}) is numerically supported by the condition
\begin{equation}\label{delta}
\delta= \min_{\Omega_0^\prime} \left\Vert \rho-\rho_{c}\right\Vert_{1}-c_{0} \ge 0,
\end{equation}
with the equality expected after minimization over all classical states $\rho_{c}$, i.e. $\Omega_0^\prime=\Omega_0$. In Fig.~\ref{fig:fignum}, we present a numerical analysis of Eq.~(\ref{delta}) through a histogram of $\delta$. This 
has been obtained for $N=10^3$ Bell-diagonal states $\rho$ randomly generated inside of the tetrahedron 
(Fig.~\ref{fig:tetraedro}). For each $\rho$, we have performed the minimization in  Eq.~(\ref{delta}) with $N_{c} = 10^6$ 
classical states $\rho_{c}$ randomly chosen from Eq.~(\ref{rho-classical}). Note that  $\delta \geq 0$, with an average value 
$\bar{\delta}=0.06$. In the inset, we have investigated the behavior of $\bar{\delta}$ as we increase the number of classical 
states $N_c$ in $\Omega_0^\prime$. For each value of $\log_{10}N_{c}$ (data point), we compute $\log_{10}\bar{\delta}$ by 
randomly selecting $10^3$ independent states $\rho$. By a linear fit (solid line) we obtain that $\bar{\delta}$ 
decreases to zero for $N_{c}\rightarrow\infty$, according to the power law 
$\bar{\delta}=0.56 \times N_{c}^{-0.16}$~\footnote{After the completion of this work, an analytical discussion 
of Eq.~(\ref{delta}) has appeared in a revised version of Ref.~\cite{Nakano:12}, which is in agreement with our numerical 
results. Moreover, it has been shown in Ref.~\cite{Ciccarello} that $D_{1}$ can be computed for generic X states.}.

{\it Monotonicity with other quantum discord measures}. Let us now apply Eq.~(\ref{1normGD}) to investigate the monotonicity of $D_1(\rho)$ with the quantum correlation measures ${\cal{Q}}(\rho)$ and $D_2(\rho)$, 
which are given by Eqs.~ (\ref{D}) and (\ref{D2}). First of all, we readily conclude that $D_{1}=0$ over the orthogonal axes $c_{1}$, $c_{2}$, and $c_{3}$, and is maximal ($D_{1}=1$) for the four Bell states, as it occurs for ${\cal Q}$ and $D_G$. Moreover, since $0 \leq c_{0} \leq 1$ and $c_{-}  \leq c_{0}$,  it follows that $c_{0}^2 \geq  \left(c_{-}^{2}+c_{0}^{2}\right)/2 \Longrightarrow D_{1}^2 \geq 2D_{G}$. From this inequality and from Eq.~(\ref{eq:INEQ}), we can find the following hierarchy for two-qubit Bell-diagonal states:
\begin{equation}
D_{1}^2 \geq 2D_{G} \geq {\cal Q}^2,{\cal N}^2.\label{DmaiorN}
\end{equation}
The inequality $D_{1} \geq {\cal N}$ that emerges from Eq.~(\ref{DmaiorN}) has also been proposed for arbitrary bipartite 
states in Ref. \cite{Debarba}, but counterexamples have subsequently pointed out in Ref. \cite{Rana:13}.

\begin{figure}[ht!]
\includegraphics[scale=0.37]{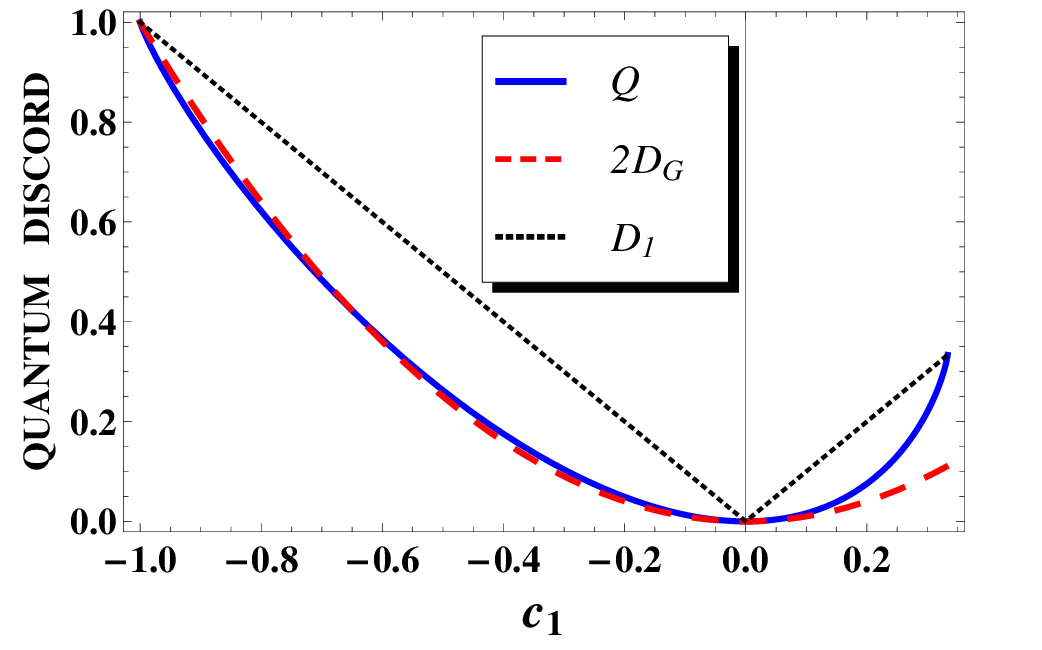} \caption{\label{fig:XXX} (Color online) Plots of ${\cal Q}$ (solid line), $2D_{G}$  (dashed line) and $D_{1}$ (dotted line) for 
SU(2)-symmetric states ($c_{1}=c_{2}=c_{3}$).}
\end{figure}
\begin{figure}
\includegraphics[scale=0.37]{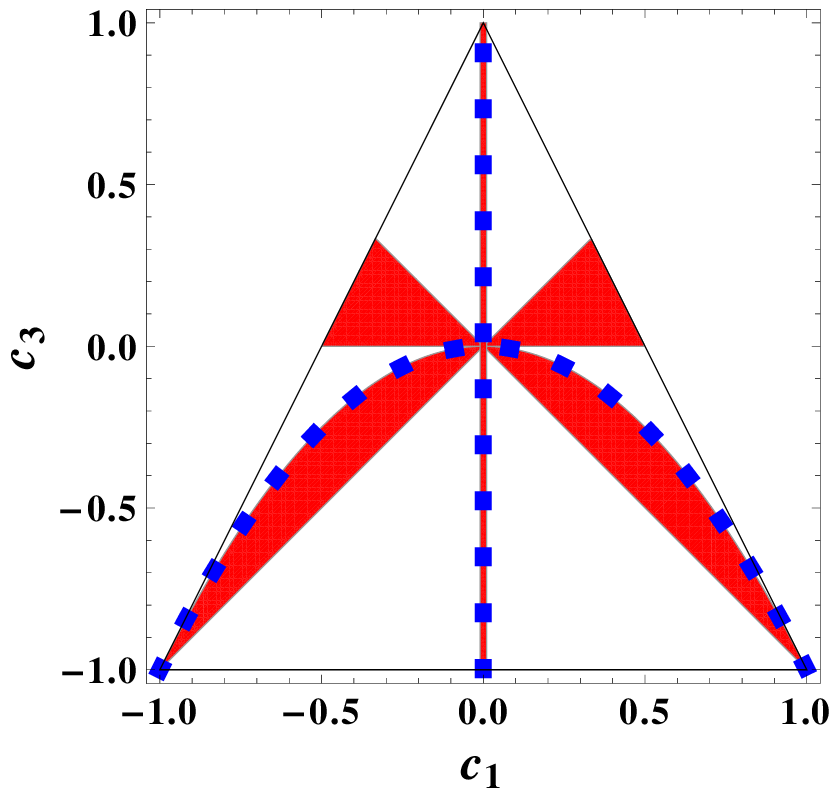} \caption{\label{fig:XXZ} (Color online) Triangle representing U(1)-symmetric states ($c_{1}=c_{2}  \ne c_{3}$). 
Shaded regions and dashed lines indicate the points for which ${\cal Q}$ is monotonically related along the $c_{3}$ direction with $D_G$ and $D_{1}$, 
respectively.}
\end{figure}

Concerning monotonicity relationships, the symmetry exhibited by the quantum state plays a fundamental role. For instance, in the case of SU(2) symmetry, i.e., 
$c_{1}=c_{2}=c_{3}$, the three measures of discord maintain the ordering of states throughout the physical region ($0\leq c_{1}\leq 1/3$), as we can observe in Fig.~\ref{fig:XXX}. 
However, this does not occur for more general classes of states. 
For instance, the triangle shown in Fig.~\ref{fig:XXZ} represents the set of physical states corresponding to 
the class of U(1)-symmetric  states, i.e.,  $c_{1}=c_{2}  \ne c_{3}$. Inside the triangle, the shaded region and the dashed lines indicate the points where ${\cal{Q}}$ 
is monotonically related along the $c_{3}$ direction with $D_{G}$ and $D_{1}$, respectively. In this situation, note that the ordering of states between ${\cal Q}$ and 
the geometric measures $D_1$ and $D_G$ is strongly violated. As the shaded region and the dashed lines do not cover the same space (a situation that occurs only when $c_{3}=-c_{1}^{2}$ and $c_{1}=0$), we also concluded that 
$D_{G}$ and $D_{1}$ are not monotonic between themselves in general.

In conclusion, the 1-norm geometric discord has by itself a conceptual importance since it is the only 
$p$-norm able to yield a well-defined quantum correlation measure. Moreover, it exhibits remarkable properties under decoherence 
for simple Bell diagonal states as, for instance, freezing and double sudden change~\cite{Montealegre:13}. 
As a future challenge, it would be useful to investigate its relevance for the advantage quantum protocols.

\vspace{0.5cm}
{\it Acknowledgments}. We thank Kavan Modi, Nivaldo Lemos, and Andreia Saguia for helpful discussions. This work is supported by the Brazilian agencies 
CNPq, CAPES, FAPERJ, and the Brazilian National Institute for Science and Technology of Quantum Information (INCT-IQ).


\end{document}